# BIS- A Blockchain-based Solution for the Insurance Industry in Smart Cities

Maedeh Sharifinejad, Ali Dorri, Javad Rezazadeh

**Abstract:** Insurance is one of the fundamental services offered to the citizens to reduce their costs and assist them in case of an emergency. One of the most important challenges in the insurance industry is to address liability challenge and the forging of documents by the involved parties, i.e., insurance company or the users, in order to increase financial gain. Conventional methods to address this challenge is significantly time consuming and costly and also suffers from lock of transparency. In this paper, we propose a blockchain-based solution for the insurance industry in smart cities (BIS). BIS creates a big umbrella that consists of the smart city managers, insurance companies, users, and sensors and devices. The users are known by changeable Public Keys (PKs) that introduces a level of anonymity. The data collected by the sensors is stored in cloud or local storage and is shared with insurance company on demand to find the liable party that in turn increases the privacy of the users. BIS enables the users to prove and share the history of their insurances with other users or insurances. Using Proof of Concept (POC) implementation we demonstrated the applicability of blockchain in insurance industry. The implementation results prove that BIS significantly reduces delay involved in insurance industry as compared with conventional insurance methods.

## Introduction

Smart city is one the emerging technologies which has received tremendous attention in recent years as a mean to reduce costs for city managers, automate city management tasks, and provide real-time services to the end users [1]. Smart city is equipped with a broad range of sensors and devices that are connected through the Internet of Things (IoT). The IoT devices can be embedded in smart vehicles or roadside infrastructures, collect data from the environment, share data with other users or Service Providers (SPs), and perform actions. One of the fundamental services offered in smart cities is insurance services which saves the user cost under particular circumstances depending on the insurance type. In the existing insurance solutions, the users sign contract with insurance companies. When the user makes claim, the insurance company sends trained investigators to investigate the situation and also may request data from other authorities, e.g., police. However, existing insurance solutions suffer from the following challenges: i) fraud detection: both insurance companies and costumers have financial benefits in conducting fraud which makes detecting fraud critical yet challenging and costly, ii) insurance history prove: the users need to purchase multiple insurances to cover a various assets which normally comes from various insurance companies. Thus, the history of the user insurance is scattered among multiple companies which makes proving the insurance history time consuming and costly, iii) Delay in finding the culprit: The time required for paying the costumers for their lost is increasing due to the difficulty in finding the liable party and gathering information for decision making by the insurance company. Mainly data gathering requires the insurance company to send experts to estimate damage. This is also costly for the insurance company that eventually impacts the insurance cost, and iv) Transparency: The costumers do not always have access to the data collected by the insurance company and cannot verify the correctness of the data and thus the decision made by the insurance company. Therefore, there is no guarantee of the accuracy of the evidence collected which limits the transparency of the liability process.

Blockchain, a distributed ledger of blocks, has received tremendous attention to serve as a decentralized framework for insurance industry due to its key features including security, anonymity, invariability and transparency. In blockchain, all transactions, i.e., communications between nodes, are verified by the participating nodes, which introduces distributed management of the blockchain and high transparency of the exchanged data. Due to its salient features, blockchain has the potential to address the outlined challenges in insurance industry in smart cities.

In this paper we propose a blockchain-based solution for insurance industry in smart cities (BIS). In BIS, the IoT devices, insurance companies, and users jointly form a public blockchain. BIS utilizes IoT devices as data sources to facilitate the liability challenge and reduce the associated cost. Additionally, the data of other participating nodes in the blockchain is employed to address the liability challenge. For example, in case of a vehicle accident, the data provided by other vehicles in the same area can be used to find the liable party. BIS enables the users to anonymously negotiate with the insurance company (of their choice) about the terms of an insurance contract. Once both sides agreed on the condition, the contract is established in the blockchain. The contract serves as genesis transaction, i.e., the first transaction in a ledger, and all corresponding transactions to the insurance are chained to the contract. Based on the contract, the insurance company request the user to either install IoT devices on premises or authorize the insurance company to access the data of existing IoT devices. The devices capture various parameters in the site and store in a cloud or local storage where the hash of the data is stored in blockchain. The latter ensures that the data cannot be tampered as in such cases the hash of the tampered data will not match with the stored hash in the blockchain.

BIS enables the smart city users to share and prove the insurance history with other insurance companies. The users receive token as evident of their insurance which can share with any other participating nodes. This also protects the privacy of the user or insurance companies as changeable PKs can be employed as the identity of any of the participating nodes.

**Related Works**

In recent years blockchain has attracted attention to enhance security and transparency in smart cities [2]. In this section, we first provide a background in blockchain and then study the existing blockchain-based solutions for insurance industry.

Blockchain is a database shared across all the participating nodes. All communications between the nodes, known as transactions, are recorded permanently in the blockchain in the form of blocks following a consensus algorithm which ensures randomness among the nodes that attempt to store transactions in the blockchain, known as miners. Each block maintains the hash of its previous block which makes blockchain immutable as altering the transactions in one block will change its corresponding hash which will be different with the hash stored in the next block. Each node can change its PK for each transaction in order to maintain its anonymity. The basic structure of a blockchain is shown in Figure 1.

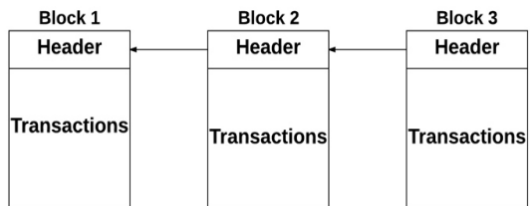

Figure 1. A high-level view of blockchain.

Ethereum [4] proposed the concept of distributed applications where the executable code of an application is stored in the blockchain and thus is run by all participants, which is also known as smart contracts. The authors in [9] proposed a blockchain-based solution that offers high transparency in vehicle insurances. The authors in [8] proposed a framework for managing vehicle insurance based on the Internet of objects and blockchain using the IoT to find some specific information about users. This information enables the insurance companies to increase its accuracy. In addition, considering the potential of the smart contract of the blockchain, determining insurance rates must be transparent. Authors in [7] proposed the use of blockchain in the business process (BPM) in smart cities.

Authors in [3] proposed a partitioned Blockchain based Framework for Auto-insurance Claims and Adjudication (B-FICA) that tracks both sensor data and entity interactions with two-sided verification. B-FICA reduces processing time, delay, and cost for practical scenarios.

Authors in [5] proposed a prototype implementation of a blockchain-based framework for automotive industry. The user agreement with insurance company along with the data of collected by the vehicle, e.g., pictures, is stored in the blockchain. The collected data is used to adjust the insurance fee of the users, e.g., a driver with dangerous driving pattern needs to pay more fee.

The existing literature in the blockchain and insurance industry suffers from the following challenges: i) Applicability: most of the existing works are only defined for smart vehicles sharing data with a single insurance company. However, in a smart city a user may have multiple assets, e.g., smart home, vehicle, etc., and thus demand multiple insurance types with different insurance companies, ii) Data collection: in some of the existing works, the data of IoT devices in multiple sites is not considered to address the liability challenge. However, this data can speed-up the process in solving the liability challenge. Additionally, in smart city multiple participants can provide data to facilitate liability. For example, in an accident, other vehicles can provide the data to find liable parties, iii) Reliance on trusted third parties (TTPs): most of the existing works are implemented in private chains, where a particular group of trusted nodes manage the blockchain. This introduces a degree of centralization and enables the central authorities to track user data and thus compromise their privacy, iv) Lack of privacy: in most of the existing works the data of the users are shared with the insurance companies which potentially introduces privacy concerns.

BIS enables the smart city users to manage all their insurances including health, vehicle, home, etc., in one place. The user privacy is maintained as the data of IoT devices are only shared with insurance companies on demand when the user made a claim. BIS is implemented on top of public blockchain that potentially increases the traceability and reduces the reliance on TTPs.

**Blockchain- based Solution for the Insurance Industry in Smart city (BIS)**

This section outlines BIS in details. The participating nodes including smart city users, insurance companies, IoT sensors, city service providers, e.g., police and city managers, jointly form a public blockchain, as shown in Figure 2, where any node can join the network without seeking permission from any authority. The participating nodes employ a changeable PK as their identity that introduces a level of anonymity. However, in some cases the users may need to be identified by the transaction recipient, e.g., an insurance company needs to identify the user that is receiving or requesting service. In such cases, the user employs key(s) known to the transaction recipient but remains anonymous to the rest of the network participants.

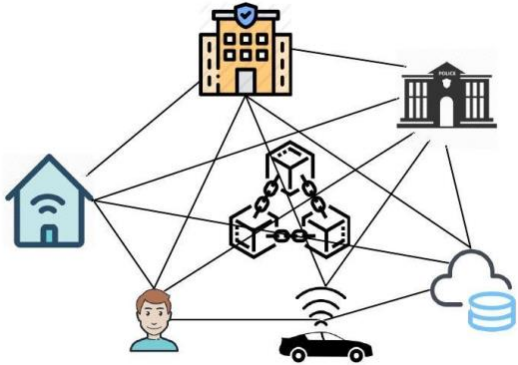

Figure 2. An overview of the BIS.

## Overview

The insurance companies store the information related to the available policies either on the chain in the metadata field of a transaction or off-chain, e.g., in the company's website. The former method enables the users to explore all available insurance companies by searching the blockchain which in turn creates a competitive environment for the smaller insurance companies by introducing high visibility for them. To reduce the associated overheads, the insurance companies store key words, e.g., insurance types, in the blockchain while storing the details off-the-chain. We assume that Distributed Time-based Consensus (DTC) algorithm is employed [6] as the consensus algorithm in BIS. The latter suits the smart city use case due to low resource consumption and self-scaling feature. DTC replaces the resource consuming consensus algorithms with time-based algorithm that demands the miners to wait for a particular time before storing new blocks. Conventional blockchains have limited throughput while DTC adjusts throughput based on the network load.

The users choose an insurance and sign contract that is stored in the blockchain in the form of a smart contract. The smart contract serves as the genesis transaction, i.e., the first transaction in a ledger, and all future transactions related to the contract are chained to the genesis transaction. To increase the accuracy of the provided services by the insurance company, reduce fraud and save time a cost in identifying the liable party, the data of multiple IoT devices is stored in a storage (either local or in a cloud) and is shared with insurance company or other authorities in case of request. Additionally, new insurance models for particular applications in smart cities demand the users to share information with the insurance company that is used to evaluate the risk and thus the insurance annual price. As an example, the car insurance company may demand driving behavior of the vehicle to identify the risk and the chance of accident by the driver. However, sharing such data compromises the privacy of the users. To address such limitations, the user stores data locally and only share on-demand. To ensure the integrity of the data, the hash of the data is stored in the blockchain. BIS enables the participants to manage their finances using blockchain system. The payments are handled in Bitcoin, or any other type of cryptocurrency.

BIS enables the users to share and prove their insurance history by introducing new tokens that are issued by the insurance companies either at the end of a contract or by user request. In summary, BIS introduces transparency and reliability in insurances process, significantly reduces cost and time, and prevents forgery in the process of determining damages and identifying the culprit at the time of the accident, which benefits the customer and insurance company.

Before signing a contract with insurance companies, the user can enter into a negotiation with the insurance company to negotiate terms or price of the insurance which is outlined in detail in follow.

## Negotiation

The user and the insurance company can negotiate over the terms or cost of a particular policy. To negotiate, the user generates a negotiation transaction (NT) which is structured as follow:

"T_ID || Insurance_PK || Condition || PK ||Sign"

Where T-ID is the identifier of the transaction, which is essentially the hash of the transaction content, Insurance_PK is the PK of the insurance company, Condition is the changes the user wish to apply, e.g., new price. In case the size of the condition is large, only the hash of it is stored in this field to ensure blockchain scalability and reduce the associated overheads. The user must maintain a local copy of the condition. In case of a dispute, the user can reveal the data and prove its not tampered with by referring to the hash in NT. Finally, the last two fields are PK and signature of the smart city user. The user and insurance company may exchange multiple NTs till reach an agreement. Once an agreement is achieved, the insurance company generates a smart contract as outlined below.

## Smart Contract (SC)

The smart contract contains all terms and conditions of the insurance. The structure of the transaction corresponding to the smart contract generation (known as SCT) is as follow:

"T_ID || User_PK || User_Sign || Insurance_PK || Insurace_Sign || Contract"

Where T_ID is as outlined above. SCT is a multi-sign transaction meaning that it requires signature of two entities to be stored in the blockchain. The next four fields are the PK and signature of the insurance company and the user. Note that the real identity of the user is known to the insurance company which makes the user anonymous only to other blockchain participants. Contract is the content of the smart contract. In case of large documents, the insurance company may store the hash of the document in the blockchain to reduce blockchain size and associated overhead. Malicious nodes may attempt to monitor the contract content to deanonymize a user or to compromise the user privacy, thus the insurance company or the user may decide to store the hash of the contract content in the blockchain.

In BIS, IoT devices are connected to the blockchain and share information to facilitate the liability challenge. The contract may request the user to install sensors on site for data collection or may request to receive data from already installed sensors in the user site, e.g., the data of a smart vehicle. Once the IoT devices are installed, the insurance company may send technicians to examine the sensor and ensure its not tampered with. The user creates a genesis transaction for each sensor which contains a virtual link to SCT by populating the T_ID of the corresponding SCT in the transaction. The IoT devices does not share data with insurance companies in real-time, instead the data is stored either in a local or cloud storage that in turn protects the privacy of the users as the data of IoT device contain privacy sensitive information captured from the everyday life of the users. The IoT devices store the corresponding hash of the data in the blockchain for integrity purpose. Similar to SC, the genesis transaction must be signed by both the smart city user and the insurance company.

By generating smart contract in the blockchain, the insurance company also records the new customer information in its server which includes policy, condition, payments, smart contract ID, PK of the IoT devices and the user. These are later used by the insurance company to retrieve details of the contract and the information of the user in blockchain. When an accident happens, the user may lodge a claim request to the insurance company using a claim request transaction as outlined below.

## Claim Request (CR)

The structure of a CR is as follow:
"T_ID || P_T_ID || Claim_Request || Data_Hash || Insurance_PK || User_PK || Sign"
Where T_ ID is the transaction identifier. P_T_ID is the identifier of the previous transaction generated by the same user which essentially chains all transactions generated by the user to the smart contract. Claim_Request contains the details of the damage the claim by the user. The user may exchange the data of the IoT devices with insurance company to facilitate the liability. In this case, the data is sent directly to the user and the hash of the exchanged data is populated in Data_Hash field. Next two fields are the PK of the insurance company and the user. Finally, the user signs the transaction and populate in Sign field.

Upon receipt of the claim request, the insurance company first verifies the user account using the User_PK field in the transaction. This ensures the claim requester has an account with the insurance company and has signed a contract. Following the insurance company verifies the signature of the user to ensure the true user has generated the request. The insurance company may need to access the data of IoT devices and thus generates a Data Access Transactions (DAT). The latter is a multisig transaction that requires the signature of the user and the insurance company to be considered as valid. The transaction is then sent to the user to be signed. The user verifies the request and grants permission to the insurance company to access the data. If the data is stored in a cloud, the insurance company uses DAT as permission to access the data. DAT authorizes the insurance company to read data from the cloud. If the data is stored locally, the user shares the data with the insurance company. The hash of the exchanged data is stored in DAT. On receipt of the data, the insurance company verifies the integrity of the received data by comparing the hash of the received data with the hash previously stored in the blockchain.

The insurance company processes the data and request further information from the user or IoT devices if needed. Third parties or technicians may also be involved to investigate the situation. In this case, the data provided by the involved parties is also shared with the insurance company and the record is stored in the blockchain chaining to the smart contract. This ensures high auditability by the end-user to ensure the type of data and the involved parties in solving the liability.

Once the insurance company made the final decision, it creates a Decision Transaction (DT) which is structured as bellow:
"T_ID || P_T_ID || Decision || User_PK || User_Sign || Insurance_PK || Insurance_Sign"
The first two fields are as discussed earlier. Decision is the final decision of the insurance company. Bothe the insurance company and the user must agree on the decision, thus the DT is a multisign transaction that requires the signature of both the user and the insurance company. Once the transaction is stored in the blockchain, the smart contract transfers the amount of claim as specified in Decision field from the insurance account to the user account.

In the event of a user's dissatisfaction with the decision of the insurance company, it can sue that in courts. Benefiting from the blockchain, the user can prove the history of its interactions with the insurance, exchanged data, and decisions made. The court verifies these data using blockchain and makes final decision. The decision is then stored in the blockchain chained to the previous smart contract.

BIS enables the users to prove their insurance history as outlined below.

## Proof of Insurance History (PIH)

One of the fundamental challenges in the existing insurance cases is to prove the history of insurance when a user decides to change its insurance provider or register for a new insurance policy. To address this challenge, we introduce proof of insurance history (PIH) token which is the signed hash of the history of the interactions of the user and the previous insurance company, i.e., the hash of the transactions chained to SCT. The structure of PIH is as below:
"Token_ID || Token_content || Metadata || User_PK || Insurance_PK || Insurance_Sign"
Where Token_ID is essentially the hash of the transaction content, Token_content is the key information about the contract which includes type of the policy, duration, and the ID of the last transaction in the blockchain generated for this contract. The latter is employed by the new insurance company to track the interactions between the user and the insurance company. Note that to protect user privacy, no data is stored in the blockchain, thus the user may decide to share the data corresponding to its history of interactions with the previous insurance. Such data is populated in Metadata field. If the size of the data is large, the data is shared off-chain and only the hash is stored in the blockchain. The PK of the user is populated in User_PK, and the last two fields contain the PK and signature of the insurance company.

The PIH enables the users to share their history of insurance, potentially to receive discount in new insurance company, while maintaining the user privacy. The new insurance company validates the data by evaluating the ledger of transactions and the data stored in the blockchain.

## Proof of Concept (PoC) Implementation

In this section we outline the details of the POC implementation of BIS. We implemented all the functions of BIS in Python and run the POC on a HP laptop (Core i5 2700 Gh, 4GB Ram). Different nodes are implemented in the laptop representing insurance companies and the users and are connected through sockets. We also defined a temperature sensor that periodically sends data to a cloud storage to be stored. We used Cloudsim [10] to represent a cloud storage and exchange data with the cloud. Cloudsim is connected to the python implementation of BIS to read/write data.

We have studied the BIS process outlined earlier in the paper including negotiation, smart contract, claim history, and PIH. The outcome of the POC proves the applicability of BIS for insurance scenarios. Using the implemented framework, we studied the performance of BIS by evaluating time and memory overheads. Time refers to the time taken for a claim request transaction to be generated and verified in the blockchain. Given that BIS is independent of the underlying consensus algorithm, we disregard consensus time for evaluations. Memory refers to the size of the transactions generated by the users. Figure 2 outlines the implementation results. During the study the user and the insurance company generate 12000 transactions. Note that the user does not store any data in the blockchain instead in the cloud using Cloudsim.

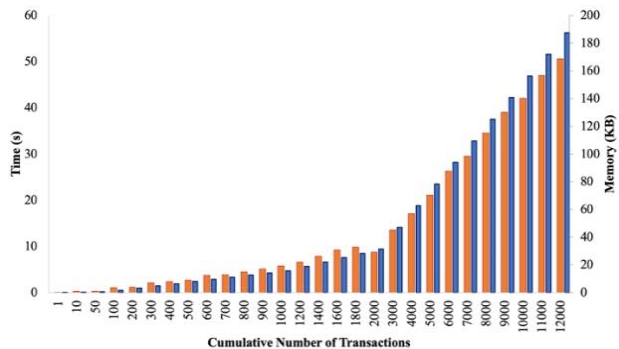

Figure 3. An evaluation of the processing time and memory consumption of BIS.

The outlined implementation results prove that BIS incurs low processing time and memory overhead. Recall that in the existing insurance infrastructures making claim and collecting data and evidence is time consuming and may demand days to complete. Additionally, the accuracy of the collected data and data trustworthiness are challenging. BIS significantly reduces the processing time as the data is shared upon receipt of the request transactions. The exchanged data can be trusted as the hash of the data is stored in the blockchain.

As outlined earlier, we have disregarded the blockchain management overhead in our evaluations. This may incur further delays and overheads on the participating nodes, e.g., the processing time to mine a block in Ethereum is around 10s. However, considering this processing time, the BIS processing time is still below the existing infrastructure which take days to process and gather data or exchange the insurance history.

## Security and Privacy Analysis

This section outlines the security and privacy of BIS. We assume that the malicious nodes may attempt to change transaction content or previously stored data and deanonymize users, however, they are not able to break the security of asymmetric encryption to cipher text data.

**Data forging:** One of the fundamental challenges in insurance industry is forging of documents or data by malicious entities, that can be either user or insurance company. BIS prevents the nodes to alter the data or communications as all interactions and the hash of the exchanged data is stored in the blockchain. Recall that blockchain is immutable ledger which makes data modification impossible.

**Privacy:** A malicious node may attempt to deanonymize a user by linking multiple transactions in the blockchain which is known as linking attack in literature [6,11]. To protect against this attack, the users employ fresh PKs to interact with multiple insurance companies or IoT devices which introduces a level of anonymity. Although we have reduced the chance of user deanonymization, linking attack is a known privacy concern in blockchain and BIS is no exception.

## Discussion and Future Research Directions

In this section, we discuss future research directions. BIS introduces a lightweight framework that includes multiple types of insurance services in a smart city. In recent years, pay as you go (PAYG) insurance models received attention from the insurance industry. Unlike conventional insurance modes where the user is charged a fixed rate annually, in PAYG insurance mode, the user is charged on the go based on his usage. For example, a smart vehicle owner is charged based on the distance he drives the vehicle monthly. However, PAYG compromises the user privacy as real-time information of the user must be exchanged with the insurances. BIS can be extended to cover PAYG requirement.

## Conclusion

In this paper, we proposed a Blockchain-based solution for insurance industry in smart cities (BIS). BIS serves as a big umbrella that contains the data of insurance companies, IoT devise, and users and facilitates the process in addressing liability challenge in case of a claim made by the users. BIS avoids fraud and denial of transactions in the insurance industry, saving time and cost and creating confidence for insurance companies and users. BIS enables the users to share the data of IoT devices with the insurance companies.

The users can also share their insurance history with blockchain participants using history tokens.

Maedeh Sharifinejad received her bachelor's degree in computer engineering from Pardis University, Iran, 2015. Currently she is a Master student in Computer Network Information Technology Engineering Dept. of Electrical and Computer Engineering, Islamic Azad University, North Tehran Branch, Tehran, Iran. Her research interest includes security and blockchain.

Ali Dorri is a Research Fellow at Queensland University of Technology (QUT), Brisbane, Australia. He received his Ph.D. from the University of New South Wales (UNSW), Sydney, Australia. He was also a Postgraduate research student at CSIRO, Australia. His research interest includes blockchain applications and challenges in adopting blockchain in large scale networks including the Internet of Things, smart cities, smart grid, and e-health. His publications in blockchain for IoT are featured among the most popular and top-cited papers in their respective venues. He has published over 25 peer-reviewed papers. His academic services also include serving on 10 conferences as a technical program committee.

Javad Rezazadeh is a researcher at the University of Technology, Sydney (UTS) in Australia and also an assistant professor at the Islamic Azad University of North Tehran Branch in Iran. He holds a Ph.D. in Wireless Communications from University Technology Malaysia (UTM) while being received Academic Excellent Student Award for the top 1 postgraduate in the year 2014. His research interests include localization systems, internet of things (IoT), Smart Cities and Machine Learning.